# KARMA Approach supporting Development Process Reconstruction in Model-based Systems Engineering

Jiawei Li, Zan Liang, Guoxin Wang, Jinzhi Lu*, Yan Yan, Shouxuan Wu, Hao Wang

**Abstract** Model reconstruction is a method used to drive the development of complex system development processes in model-based systems engineering. Currently, during the iterative design process of a system, there is a lack of an effective method to manage changes in development requirements, such as development cycle requirements and cost requirements, and to realize the reconstruction of the system development process model. To address these issues, this paper proposes a model reconstruction method to support the development process model. Firstly, the KARMA language, based on the GOPPRR-E meta-modeling method, is utilized to uniformly formalize the process models constructed based on different modeling languages. Secondly, a model reconstruction framework is introduced. This framework takes a structured development requirements based natural language as input, employs natural language processing techniques to analyze the development requirements text, and extracts structural and optimization constraint information. Then, after structural reorganization and algorithm optimization, a development process model that meets the development requirements is obtained. Finally, as a case study, the development process of the aircraft onboard maintenance system is reconstructed. The results demonstrate that this method can significantly enhance the design efficiency of the development process.

Jiawei Li
Yangtze Delta Region Academy of Beijing Institute of Technology
Jiaxing 314019, Zhejiang, China
lijiawei6866@163.com

Guoxin Wang · Yan Yan · Shouxuan Wu
Beijing Institute of Technology
Beijing 100081, China
wangguoxin@bit.edu.cn , yanyan331@bit.edu.cn , shouxuanwu.bit@gmail.com

Zan Liang
Beijing Institute of Astronautical System Engineering
Beijing 100076, China
liangzan115@126.com

Jinzhi Lu
Beijing University of Aeronautics and Astronautics
Beijing 100191, China
jinzhl@buaa.edu.cn

Hao Wang
COMAC Beijing Aircraft Technology Research Institute
Beijing 102211, China
wanghao8@comac.cc

Article sample

# 1 Introduction

The development process consists of design activities, playing a crucial role in complex systems development. It provides an overarching framework and guidance for system development, encompassing various stages, activities, and tasks. It ensures the orderly progression of development work and can enhance the efficiency and quality of development outcomes. As system complexity continues to rise, the system development process is characterized by interdisciplinary collaboration and intricate relational coupling[1].

Model-based systems engineering (MBSE) offers a methodology for addressing the complexities inherent in the development of complex product systems. Within MBSE, the development process can be depicted and expressed through modeling techniques, allowing for graphical representation that facilitates intuitive and clear description of the development process. During the iterative design process of a system, design activities are retained and inherited. Typically, the majority of process elements from the original process model can be utilized, with only minor adjustments made to expedite the design of a newly developed process. Currently, this process is often manually executed by engineers. While it does improve efficiency to some extent compared to rebuilding the development process model entirely, it still demands a significant amount of time and presents challenges in ensuring that the newly constructed development process model aligns with the development requirements. Therefore, there is a pressing need for an automated or semi-automated method to streamline this process.

Model reconstruction provides a method to address the challenges mentioned above. Model reconstruction involves optimizing the structure and content of the original model to obtain a new model in response to changes in development requirements. It is particularly suitable for the iterative design process of products, as it not only meets product requirements but also enhances the efficiency and quality of model construction in the development process. However, there are several challenges associated with the reconstruction of the development process model: 1) The reconstruction of the development process model can be viewed as an optimization problem, where demand constraints often exist in natural language form. However, existing model reconstruction algorithms utilize mathematical models and representation of reconstruction constraints, making it challenging to directly apply demand constraints expressed in natural language to these algorithms. 2) The development process can be expressed using various data specifications such as SysML activity diagrams and BPMN flow charts. However, process models with different data specifications exhibit heterogeneous semantics and syntax. Consequently, it is difficult for current model reconstruction methods to achieve reconstruction of development process models with multiple data specifications. 3) There is a lack of standardized semantic specifications to support the expression of complex topological interrelationships in the optimization process of the development process model. Therefore, there is a pressing need for computer-driven automatic reconstruction of the development process model to address these challenges.

To address the aforementioned challenges, this paper introduces a framework aimed at supporting the reconstruction of complex heterogeneous development process models. The proposed method leverages the KARMA language, based on the GOPPRR-E meta-modeling method, to uniformly describe development process models across different domains and to formalize the expression of the optimization problem space. Additionally, it proposes standardized semantic specifications to facilitate the description of reconstruction requirements and employs natural language processing methods for the processing and analysis of demand constraints. Furthermore, the framework utilizes the branch-and-bound algorithm to facilitate the optimization process of model reconstruction.

The rest of the document is arranged as follows. Section 2 introduces related work on natural language processing and model reconstruction. Section 3 proposes a framework based on the GOPPRR-E meta-modeling method for development process model reconstruction, and then Section 4 describes a case of aircraft onboard maintenance system development process reconstruction to verify the proposed method. Finally, we provide conclusions in Section 5.

## 2 Related Work

As input to the development process model reconstruction framework, demand constraints often manifest in natural language form. The analysis of demand constraints necessitates the utilization of Natural Language Processing (NLP) technology. NLP is a subfield of artificial intelligence concerned with the computational processing and comprehension of human language[2]. NLP researchers have combined neural networks with deep learning to achieve forefront advancements in various tasks including chatbot development[3], machine translation[4], and numerous others. However, NLP methods relying on deep learning or machine learning typically demand a substantial amount of annotated data for effective model training, which may not be readily available in the domain of development process model reconstruction. Moreover, the accuracy and domain specificity of these methods for natural language processing often fall short, posing challenges in meeting the contextual and requirement needs of specific domains in the reconstruction of development process models. In contrast, rule-based information extraction techniques offer notable advantages in these aspects.

Rule-based information extraction technology indeed offers a promising solution to the aforementioned challenges. By tailoring domain-specific knowledge extraction rules, this approach can readily adapt to the requirements of specific fields or tasks. In the model-based systems engineering domain, Jahan M[5] and others proposed a method to automatically generate UML sequence diagrams. The input text should follow certain constraints, and then be processed into simple statements, and then the sequence diagrams are generated according to the rules. Chen J[6] and others proposed a method for automatically generating SysML models based on Chinese text requirements. This method involves constructing domain dictionaries, extracting rules, segmenting and annotating requirement texts, and ultimately transforming requirement elements into SysML requirement diagrams. However, research on the reconstruction of development process models within the model-based systems engineering domain is currently relatively limited. Consequently, there is also a scarcity of domain dictionaries and knowledge extraction rules applicable to the reconstruction of development process models, thereby hindering the utilization of rule-based information extraction techniques for extracting information on the requirements of development process model reconstruction.

In model-based systems engineering, the development process model can be expressed by a variety of data specifications, such as SysML[7], BPMN[8], and UML[9]. However, these data specifications are heterogeneous, and a modeling language is needed to uniformly express these heterogeneous data specifications. Lu et al. [9]proposed a text modeling language KARMA based on the GOPPRR-E meta-modeling method to achieve the unified formalization of multi-domain system models. However, KARMA lacks the ability to describe the development process model reconstruction currently.

## 3 A model reconstruction method for development process model

### *3.1 Formalization of development process model based on GOPPRRE*

In order to realize the construction of multi-data specification development process models, this paper uses the KARMA language on the basis of the GOPPRR-E meta-modeling method to unify and to formalize the heterogeneous process models. The MOF framework[11] is the basis of the GOPPRR-E meta-modelling method.

As shown in Figure 1, the MOF consists of four layers: M3-M0. The M3 layer refers to the meta-meta model, which is the basic elements that constitute the meta-model. The M2 layer refers to the metamodel layer, which serves as an instance of the meta-meta model in the M3 layer and supports the definition of the model in the M1 layer. Modeling languages are defined based on metamodels. The M1 layer represents the model layer. Models serve as instances of metamodels and can support the development of complex systems. The M0 layer represents the perspective of the real system, abstractly described by the model in the M1 layer.

The M3 layer in the GOPPRR-E meta-modeling method contains six meta-meta-models: graph, object, point, relationship, attribute and role. A graph is a collection of objects and their associated relationships. An object refers to an entity in the graph. A point represents a port in an object and must exist attached to the object. Relationships are used to represent the interrelationship between objects. The beginning and end of the relationship are connected to the objects through roles. Roles are bound to objects or points attached to objects. Properties refer to the properties of the other five metamodels (graph, object, relationship, point, and role).

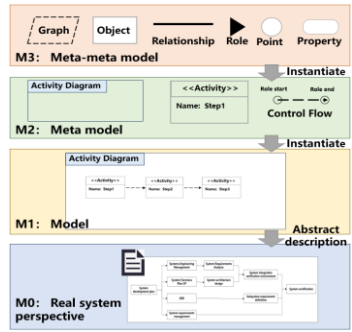 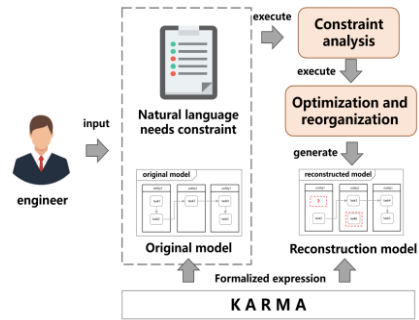

**Fig. 1.** The GOPPRR-E meta-modeling approach based on MOF

**Fig. 2.** The Model Reconstruction Framework

Based on the GOPPRR-E meta-modeling approach, a semantic modeling language named KARMA[9] has been developed, supporting the unified representation of different modeling languages such as SysML, BPMN, and UML. The underlying data specifications of these modeling languages are diverse. Additionally, development process models can be articulated using these modeling languages. As a result, the KARMA language enables the unified representation of heterogeneous development process models.

### 3.2 Developing a process model reconstruction framework

Building upon KARMA for the uniform expression of heterogeneous process models, we present a model reconstruction framework tailored for KARMA process models. Initially, the model reconstruction process for the KARMA model is outlined. Following this, a methodology for implementing the model reconstruction process is proposed. The overall framework is depicted in Figure 2. Engineers articulate the requirement constraints for model reconstruction in natural language and prepare the model accordingly. Subsequently, constraint analysis and optimization restructuring processes are undertaken, culminating in the reconstructed model.

(1) constraint analysis

| G | E | G | E |
|---|---|---|---|
| should、must | shall | less than、lower than | less |
| add、Increase、execute | add | maximum、greatest、most | max |
| reserve、save、retain | reserve | minimum、least | min |
| property | property | task、activity | task |
| more than、higher than | greater | | |

**Fig. 3.** User Dictionary (Part)

| Template Type | Template Content |
|---|---|
| Selection Entity Template | <the new model> <shall> <contain> {entity1}、{entity2} …… |
| Augmented task Template | {entity} <shall><add> {task} |
| Retained task Template | <reserve>{property} <greater/less> {value} <task> |
| Objective Function Template | {property} <max/min> |
| Constraint Template | {property} <greater/less> {value} |

**Fig. 4.** Template Type and Content

Constraint analysis involves processing the input natural language development requirements text. It employs a rule-based development requirements text knowledge extraction method to capture and store the extracted constraint information in a standardized format. This extracted information serves as input for the optimization and reorganization phase. The development requirements span across five categories: selection entity information, augmented task information, objective function information, retained task information, and constraint information.

As shown in Figure 2, Knowledge extraction refers to text processing technology that extracts specified types of entities, attributes and other information from natural language text and forms structured data. The first step in the rule-based demand constraint text knowledge extraction method is to define a custom user dictionary. This article uses the JIEBA library[12] as the dictionary. However, relying solely on the words that come with the JIEBA library cannot meet the need to extract constraints. Therefore, we utilize the custom dictionary feature of the JIEBA toolkit, the rules for custom dictionaries include two parts, as follow:

$$G = (V, E) \qquad (1)$$

Among these categories, G represents a customized user dictionary, while V represents entities expected to be extracted through named entity recognition. E denotes parts of speech, typically including nouns (n), proper nouns (nz), among others. These parts of speech can also be customized to fulfill the requirements of knowledge extraction.

For the model reconstruction framework, this article establishes a series of customized user dictionaries tailored for model reconstruction. As shown in Figure 3, several dictionaries are delineated as follows.

In the second step of the model reconstruction framework, semantic specifications of the natural language requirement constraint knowledge extraction template are defined, leveraging the custom user dictionary. The development of knowledge extraction rule templates is distinctly categorized into the following five types. As shown in Figure 4.

Article sample

| Constraint Information Type | Constraint Information Content |
|---|---|
| Selection entity constraint information (represented by ESC) | ESC={entity1, entity2, ⋯} |
| Augmented task constraint information (represented by AAC) | ESC={entity1, entity2, ⋯} |
| Retained task constraint information (represented by ARC) | ARC={property, greater/less, target value} |
| Objective function constraint information(represented by TFC) | TFC={property, max/min} |
| Constraint information(represented by CC) | CC={property, greater/less, target value} |

**Fig. 5.** Constraint Information Type and Content

Among these, the entity selection template delineates the specification for expressing entity selection information. In SysML and BPMN, swim lanes are used to represent different participants, roles or departments in the process model to help clearly demonstrate the collaboration and division of labor of the process[13]. In the new process model, certain entities and their associated tasks may become unnecessary. Therefore, it is essential to select the entities required for the new process model. The augmented task template outlines the specifications for expressing augmented task information. In comparison to the original process model, the new process model may require the addition of new tasks to meet the new system requirements. The retained task template specifies how retained task information is expressed. Throughout the model reconstruction process, certain tasks from the original process model may remain essential. Therefore, it is imperative to retain these tasks and exclude them from the subsequent linear optimization process. The objective function template delineates the specifications for expressing objective function information. In defining the objective function of a linear optimization problem, the objective function category serves as the attribute within the activity, encompassing factors such as time, cost, and importance. The value of the objective function is derived from the summation of attribute values across all tasks within the process model. The constraint template delineates the specification for expressing constraint information. We should define the constraints of a linear optimization problem, where the category of the constraints is the attributes in the task, such as time, cost, and importance, and its value is the sum of the values of an attribute in all objects in the process model.

The third step involves information extraction. Initially, tokenization is conducted, followed by part-of-speech tagging on the development requirements text, with the tokenized results being outputted. Subsequently, the tokenized results undergo further enrichment of semantic information through part-of-speech tagging. Information extraction is then performed utilizing the regular expression chunker from the NLTK library, whereby the knowledge extraction rule templates defined in the second step are mapped to NLTK chunking templates. Through this process, tokenized results are chunked, facilitating the extraction of meaningful information blocks from the textual data. Finally, the required text requirement constraint information is extracted for utilization in the subsequent optimization and reorganization phase. As shown in Figure 5, the Constraint Information Type and Content are illustrated.

(2) Optimization and reorganization

The optimization and reorganization process of the KARMA process model can be segmented into two stages: structural reorganization and algorithm optimization. Structural reorganization involves the preliminary restructuring of the process model based on demand constraint information. Algorithmic optimization utilizes algorithms to optimize the structure of the process model in alignment with the new system requirements.

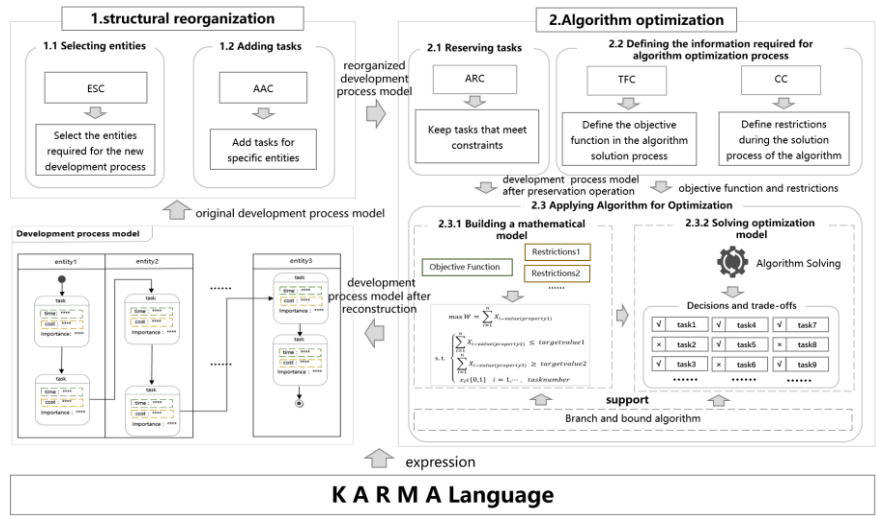

**Fig. 6.** The process of optimization and reorganization

1) Structual reorganization

Taking the original model as input, the structural reorganization operation is executed. The structural reorganization phase entails two steps: selecting entities and adding tasks.

**Selecting entities:** As shown in 1.1 of Figure 6, based on the extracted constraint information (ESC), entities required for the new process model are selected from the original process model.

**Adding tasks:** As shown in 1.2 of Figure 6, based on the extracted constraint information AAC, tasks are added to the specified entities, which will meet the requirements of the new development process.

2) Algorithm optimization

Taking the restructured process model as input, the algorithm optimization operation is executed. The algorithm optimization stage comprises three steps: reserving tasks, defining the information required for the optimization process, and applying the Algorithm for Optimization.

**Reserving tasks:** As shown in 2.1 of Figure 6, based on the extracted constraint information ARC, tasks that meet the conditions are retained to avoid being included in the subsequent optimization process. Then, input the development process model after the reservation operation into step 2.3.

**Defining the information required for algorithm optimization process:** As mentioned above, the objective function and constraints in the algorithm optimization process are defined based on the sum of the specified attributes of all tasks in the process model. Therefore, as shown in 2.2 of Figure 6, the objective function information TFC is input into the algorithm optimization process as the objective function.The condition information CC is input into the algorithm optimization process as a constraint condition.Then, input the objective function and restrictions into step 2.3.Once the objective function and constraint conditions are input into the algorithm optimization process, an appropriate algorithm must be selected to solve the optimization problem. Given that most variables in the development process model are integers, and the general constraints are linear constraints, the integer programming algorithm is chosen. The branch and bound algorithm is a powerful algorithm for solving combinatorial optimization problems, as it divides the problem space into multiple sub-problems and gradually eliminates branches that are unlikely to produce optimal solutions, thereby efficiently searching for optimal solutions[14].

**Applying Algorithm for Optimization:** In the model reconstruction framework proposed in this article, the algorithm optimization part mainly focuses on linear programming, so the branch and bound method is used in the algorithm optimization process of the development process model. As shown in 2.3 of Figure 6, the specific optimization process is as follows:

● Building a mathematical model: As shown in 2.3.1 of Figure 6, according to the required attribute items in TFC and CC, extract the attribute values corresponding to each task in the process model, and convert them into objective function expressions and constraint expressions. Take the following mathematical model as an example, Objective function formula (2) represents the maximum value of the sum of the values of attribute 1 of all tasks in the process model; constraint formula (3) represents that the sum of the values of attribute 2 of all tasks in the process model is less than or equal to the target value 1; constraint formula (4) indicates that the sum of the values of attribute 3 of all tasks in the process model is greater than or equal to the target value 2; constraint (5) indicates whether to retain the task, $x_i = 0$ means discarding the task, $x_i = 1$ means retaining the task Task.

$$\max W = \sum_{i=1}^{n} X_{i-value(property1)} \quad (2)$$

$$\text{s.t.} \begin{cases} \sum_{i=1}^{n} X_{i-value(property2)} \leq targetvalue1 & (3) \\ \sum_{i=1}^{n} X_{i-value(property3)} \geq targetvalue2 & (4) \\ x_i \in \{0,1\} \quad i = 1,\cdots, tasknumber & (5) \end{cases}$$

- Solving the optimization model: As shown in 2.3.2 of Figure 6, we employ the branch and bound algorithm to solve the aforementioned mathematical model, balancing each task in the process model, and ultimately obtaining the optimal solution for the process model scheme.

After optimizing and reorganizing the original process model, a model solution that meets the demand constraints is obtained and output to the MetaGraph2.0 tool. Then, the entire model reconstruction process is completed.

## 4  Case Study

In this section, we conduct a study on the development process of the aircraft onboard maintenance system to validate the usability of the proposed method. The aircraft onboard maintenance system aims to enhance the reliability of aircraft systems through real-time monitoring, fault diagnosis, and remote maintenance. The development process of the old aircraft's airborne maintenance system involves five entities, namely suppliers, airborne health management professionals, and others. It also comprises 49 development process steps, such as scenario analysis of the airborne maintenance system. The process of developing the new aircraft's onboard maintenance system involves participation from suppliers, airborne health management professionals, member management specialists, test and validation experts, and customer service companies. Additionally, suppliers are required to perform the "Airborne Maintenance and Health Management System RFI Response" task. The total development time for the aircraft's onboard maintenance system does not exceed 2500 hours, and the total development cost does not exceed 15,000,000. The goal of model reconstruction is to prioritize steps based on their overall importance, and steps with an importance level above 90 should be retained.

As shown in Figure 7-A, a graphical activity flow model (ACT) is constructed through KARMA language to describe the specific development process of the aircraft onboard maintenance system. As shown in Figure 8, different GOPPRR meta-models are used in ACT to express different concepts. For example, the concept of Process can be described by the Object metamodel, which can be used as the type of Object instance and represented as "Object [Process]" in KARMA. The Process object has three attributes: time, cost and importance.

The development process of the aircraft onboard maintenance system is reconstructed as follows. Firstly, as shown in Figure 7-B, we input the development requirements, including selection entity information, augmented task information, objective function information, retained task information, and constraint information. After program processing, the required constraint information is generated, including ESC,AAC,ARC, TFC and CC. Based on the extracted constraint information, the final model solution is obtained after reorganizing structure, optimizing algorithm and selecting tasks in the process. As

Article sample

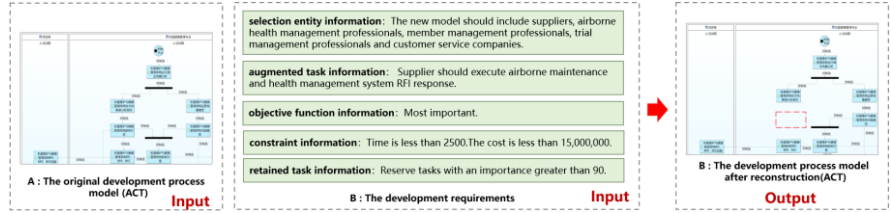

**Fig. 7.** The reconstruction process of the development process model for the aircraft's onboard maintenance system

**Fig. 8.** Elements of KARMA to express concepts of ACT

**Fig. 9.** Comparison between manual model building and model reconstruction

shown in Figure 7-C. This article contrasts the manual construction of the process model with the construction of the process model through model reconstruction. The specific results are shown in the Figure 9. Analyzing three sets of experimental data, the results indicate that using the model reconstruction method saves 74.6%, 75.4%, and 73.8% of time compared to manual methods, respectively.

## 5 Conclusion

This paper presents a method for formalizing and reconstructing multi-domain process models. It employs the KARMA language, based on the GOPPRR-E meta-modeling method, to uniformly express heterogeneous process models. Building upon this, a model reconstruction framework is introduced to facilitate the iterative design of process models. Finally, the development process of the aircraft onboard maintenance system was reconstructed based on new requirements, thus validating the effectiveness of the proposed method. In the future, the constraint analysis aspect of the model reconstruction framework will likely transition to more sophisticated natural language processing methods. Complex deep learning models, for instance, can enhance the comprehension of demand constraint information, enabling the efficient and accurate processing of constraint information embedded in natural language.

Article sample

# Acknowledgement

This work was supported by The National Ministry project (No. D020101 and No. JCKY2021203A001).